\documentclass[runningheads]{llncs}
\usepackage[labelfont=bf]{caption}
\usepackage{graphicx}
\usepackage{amsmath}
\usepackage{optidef}
\usepackage[export]{adjustbox}
\usepackage{tabularray}

\usepackage{color}

\begin{document}

\title{Bayesian optimization of the layout of wind farms with a high-fidelity surrogate model}
\titlerunning{Bayesian optimization of the layout of wind farms}
\author{Nikolaos Bempedelis\inst{1} \and Luca Magri\inst{1,2}}
\authorrunning{N. Bempedelis and L. Magri}
\institute{
Department of Aeronautics, Imperial College London, London SW7 2AZ, UK \and
The Alan Turing Institute, London NW1 2DB, UK
}

\maketitle              

\begin{abstract}
We introduce a gradient-free data-driven framework for optimizing the power output of a wind farm based on a Bayesian approach and large-eddy simulations. In contrast with conventional wind farm layout optimization strategies, which make use of simple wake models, the proposed framework accounts for complex flow phenomena such as wake meandering, local speed-ups and the interaction of the wind turbines with the atmospheric flow. The capabilities of the framework are demonstrated for the case of a small wind farm consisting of five wind turbines. It is shown that it can find optimal designs within a few iterations, while leveraging the above phenomena to deliver increased wind farm performance.

\end{abstract}

\section{Introduction}

Today, the need for renewable energy sources is more urgent than ever. In the UK, wind power is the largest source of renewable electricity, and the UK government has committed to a further major expansion in capacity by 2030. However, currently installed wind farms do not produce as much power as expected because the majority of the turbines operate within the wake field of other turbines in the farm. A wind turbine operating within a wake field is an issue for two reasons. First, the reduction of its power output due to wind speed deceleration, and, second, the increase of fatigue loads due to increased wind fluctuations.

Wake effects can be minimised by optimally arranging the wind turbines over the available land. Typically, wind farm layout optimization (WFLO) is carried out with low-fidelity flow solvers (wake models) as objective function (farm power output) evaluators \cite{shakoor2016wake,porte2020wind}. Wake models are based on simplified assumptions for the wakes of porous disks, and do not account for several mechanisms including unsteadiness, non-linear interactions, or blockage, to name a few. As a result, optimization based on wake models misses out on a number of opportunities for performance gains through manipulation and, possibly, exploitation of these phenomena. Furthermore, wake models typically provide discontinuous solutions, which renders their use within gradient-based optimization algorithms problematic. Nevertheless, wake models are almost invariably used in layout optimization studies due to their low computational cost (a single evaluation typically runs in under a second).

More accurate approximations of the wind farm flow have been considered in only a limited number of works \cite{king2017optimization,antonini2018continuous}. In these works, the wind farm layout was optimized by an adjoint approach and steady-state Reynolds-averaged Navier-Stokes (RANS) simulations. The use of steady RANS allowed capturing a number of the aforementioned phenomena. Nevertheless, the assumption of a steady flow means that the wake dynamics and the wake-to-wake and atmosphere-to-wake interactions, which are critical for the wind farm layout optimization problem, were not appropriately accounted for.    

In this work, we present a gradient-free framework for optimizing the output of a wind farm based on a Bayesian approach and high-fidelity large-eddy simulations of the flow around the wind farm.  Bayesian optimization is a suitable optimization strategy due to the  multi-modality of the WFLO problem (which makes gradient-based methods prone to getting stuck in local extrema), and the high cost of evaluating the objective function (at least when an accurate model of the flow field is desired, as in our study). The structure of the paper is as follows. The data-driven optimization framework is described in section \ref{sec:methodology}. Section \ref{sec:results} discusses its application to a  wind farm layout optimization problem. Finally, section \ref{sec:conclusions} summarises the present study.

\section{Methodology}
\label{sec:methodology}

\subsection{The optimization problem}
We aim to maximise the overall power output $P$ from $N$ different wind turbines experiencing $K$ different wind states by controlling their position $\boldsymbol{c} = \left[\boldsymbol{x}, \boldsymbol{y} \right]^T$, with $\boldsymbol{x} = \left(x_1, \dots, x_N \right)$ and $\boldsymbol{y} = \left(y_1, \dots, y_N \right)$, within a given space $\boldsymbol{X}$. The space $\boldsymbol{X}$ corresponds to the available land where the wind turbines may be installed. To avoid overlap between the different wind turbines, we enforce a constraint that ensures that their centers (i.e. their position) are spaced at least one turbine diameter $D$ apart. The optimization problem can be expressed as

\begin{argmaxi}|s|
{\boldsymbol{c}}
{\sum_{n=1}^{N} \sum_{k=1}^{K} a_k P_{n,k} }
{}{}
\addConstraint{\boldsymbol{c} \in \boldsymbol{X}}
\addConstraint{ ||\boldsymbol{c}_i - \boldsymbol{c}_j|| > D, \textrm{ for } i,j=1,\dots,N \textrm{ and } i \ne j}{}
\end{argmaxi}

with $a_k$ being the weight (i.e. probability) of each wind state, which are obtained from the local meteorological data (obtained via a measurement mast).

\subsection{Flow solver}
The power output of the wind turbines is computed with the open-source finite-difference framework \texttt{Xcompact3D} \cite{bartholomew2020xcompact3d}, which solves the incompressible filtered Navier-Stokes equations on a Cartesian mesh using sixth-order compact schemes and a third-order Adams-Bashforth method for time advancement \cite{LaizetLamballais2009}. Parallelisation is achieved with the \texttt{2Decomp \& FFT} library, which implements a 2D pencil decomposition of the computational domain \cite{LaizetLi2011}. The wind turbines are modelled with the actuator disk method. The Smagorinsky model is used to model the effects of the unresolved fluid motions. Finally, to realistically model the interaction of the wind farm with the atmospheric flow, a precursor simulation of a fully-developed neutral atmospheric boundary layer is performed to generate the inlet conditions. For more details on the numerical solver and a related validation study, the reader is referred to \cite{bempedelis2023turbulent}. 

\subsection{Bayesian optimization algorithm}
Bayesian optimization is a gradient-free optimization technique that consists of two main steps. First, a surrogate model (here a Gaussian Process) of the objective function is computed given knowledge of its value for a set of parameters. This also quantifies the uncertainty of the approximation. Second, it proposes points in the search space where sampling is likely to yield an improvement. This is achieved by minimising an acquisition function. In this work, we make use of  the \texttt{GPyOpt} library \cite{gpyopt2016}. In particular, we use the Mat\'ern 5/2 kernel and the Expected Improvement acquisition function, which is minimised using the L-BFGS algorithm. 
Bayesian optimization requires a number of initial samples to  start. In this work, these are obtained using both the Latin hypercube sampling technique and a custom function that targets a uniform distribution of the wind turbines over the available land $\boldsymbol{X}$ whilst favouring placement on the domain boundaries (the initial layouts used in this study are shown in figure \ref{fig:opt_vis}). For a more thorough description of Bayesian optimization, along with an example of it being used to optimize a chaotic fluid-mechanical system, the reader is referred to \cite{huhn2022gradient}.

\section{Results}
\label{sec:results}

\begin{figure}
\includegraphics[width=\textwidth]{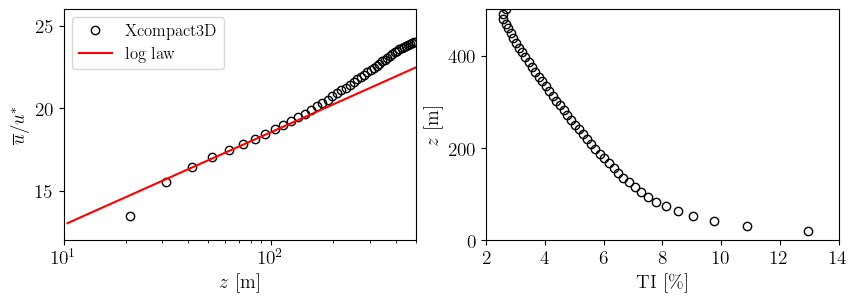}
\caption{Mean velocity (left) and turbulence intensity (right) of the simulated atmospheric boundary layer.} 
\label{fig:ABL_data}
\end{figure}

The data-driven optimization framework is deployed on the following problem. The available land is a square of size $6D \times 6D$, where $D=100$ m is the turbine diameter. The wind blows from a single direction (easterly) and at constant speed.  The atmospheric boundary layer is characterised by friction velocity $u^* = 0.442$ m/s, height $\delta=501$ m and roughness length $z_0 = 0.05$ m, which correspond to conditions in the North Sea \cite{wu2015modeling}. The velocity at the hub height of the turbines, $h=100$ m, is $U_{h} = 8.2$ m/s and the turbulence intensity at the same level is $TI_h = 7.4\%$ (see figure \ref{fig:ABL_data}). The wind turbines operate at a thrust coefficient $C_T=0.75$.
The size of the computational domain is $2004 \times 1336 \times 501$ m in the streamwise, spanwise, and vertical directions. It is discretised with $193 \times 128 \times 49$ points, respectively  (amounting to $\approx 5\times 10^6$ degrees of freedom at each time step). The land where the turbines can be placed starts at $x=3.68D$ from the upstream boundary. Periodic conditions are enforced on the lateral domain boundaries.  A time step $\Delta t = 0.15$ s is used, with the maximum CFL number remaining under $0.18$.  Statistics are averaged over a one-hour time period following one hour of initialisation  (each period corresponds to $\approx 15$ flow-through times based on the hub-height velocity). 

\begin{figure}
\centering
\includegraphics[width=.49\textwidth]{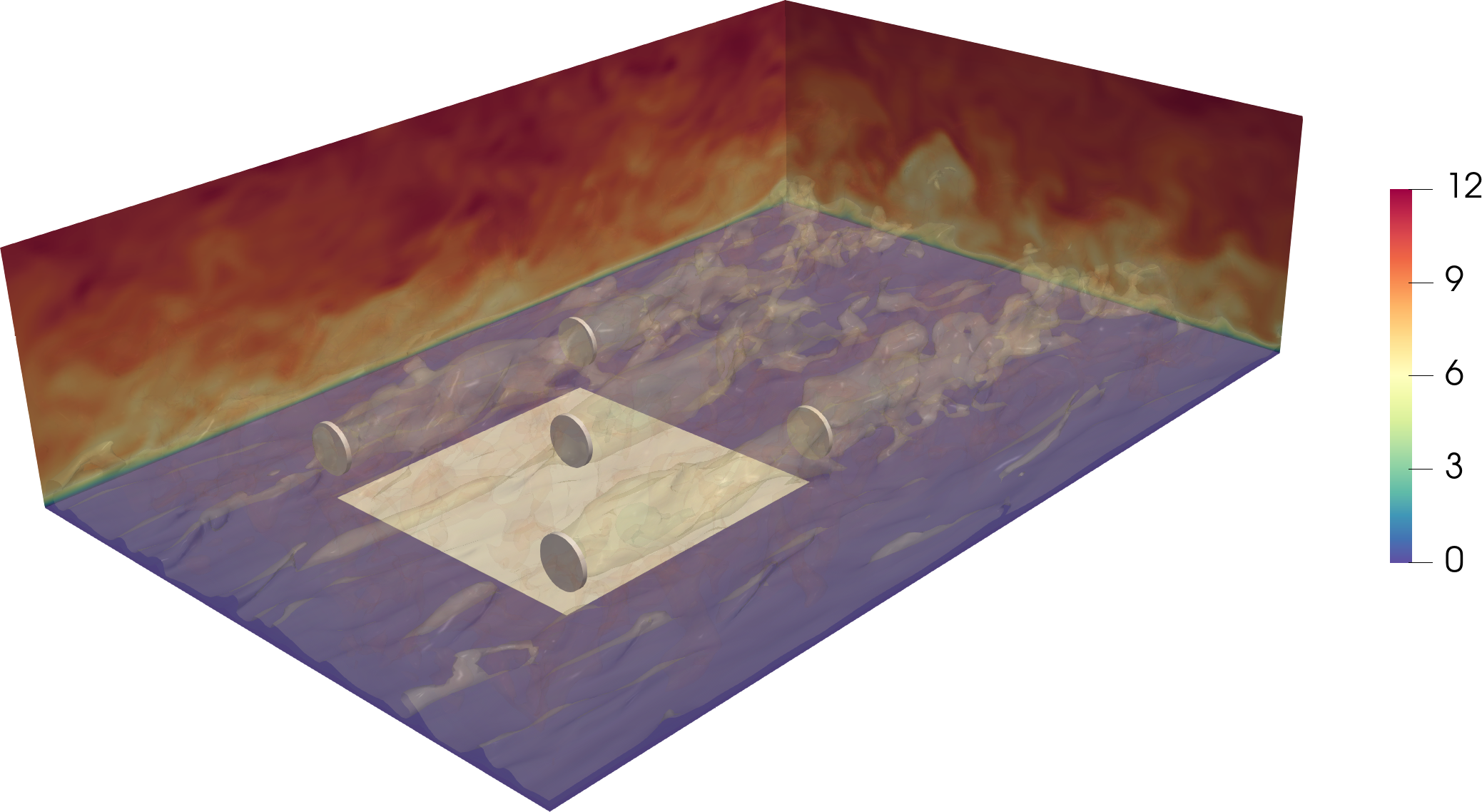}
\includegraphics[width=.47\textwidth]{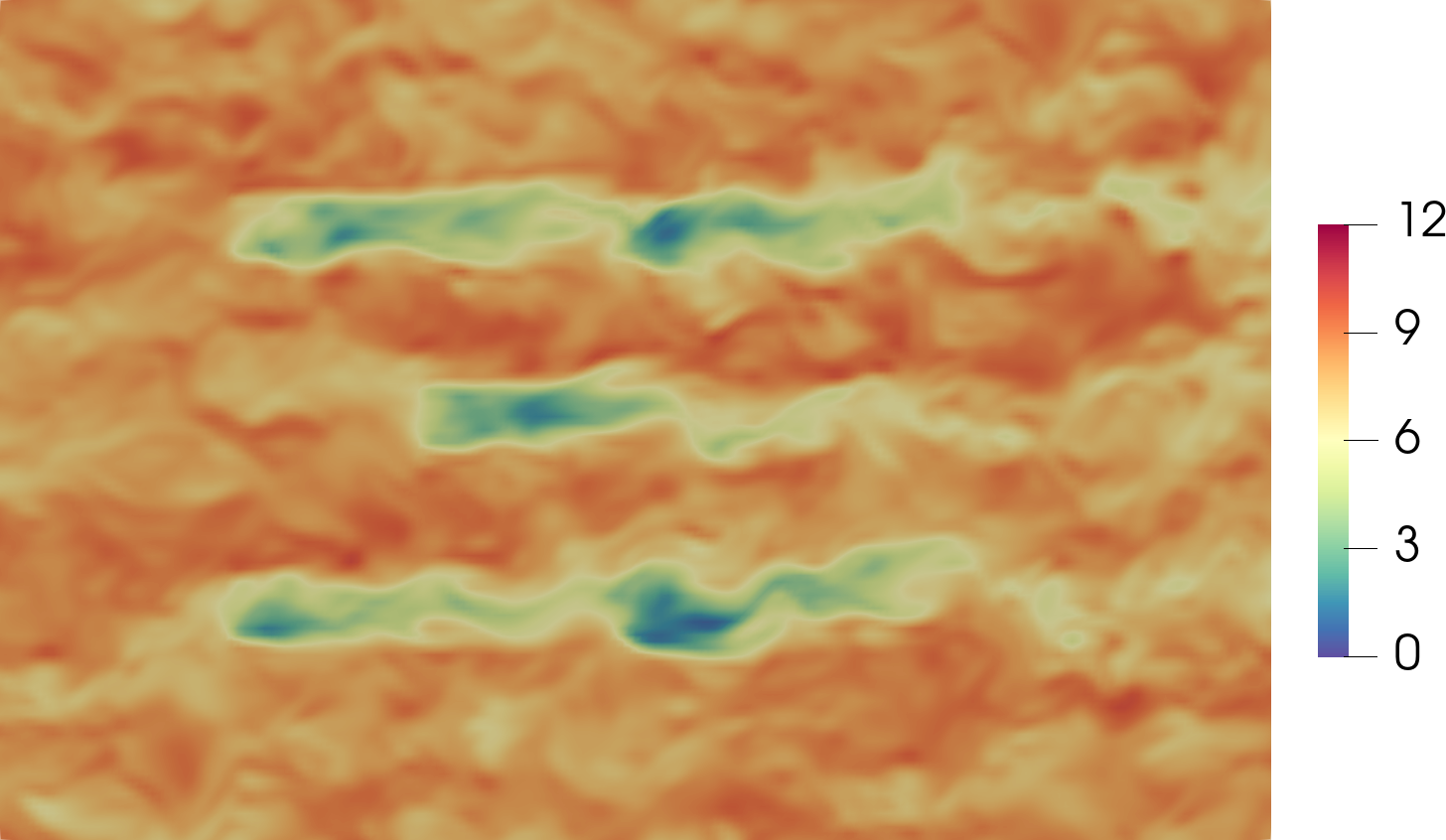}
\caption{ Instantaneous streamwise velocity for layout \#5. Three-dimensional view including a square indicating the available land and five disks indicating the turbine rotors (left). Horizontal cut at the turbine hub height (right).}
\label{fig:ABL_vis}
\end{figure}

We consider a wind farm consisting of five wind turbines. Five layouts are used to initialise the Bayesian optimization. Four are generated with the Latin hypercube sampling technique and one with the custom function (see section \ref{sec:methodology}). A snapshot of the streamwise velocity in the latter case is presented in figure \ref{fig:ABL_vis}, which shows the turbulent nature of the flow field.

\begin{figure}
\includegraphics[width=\textwidth]{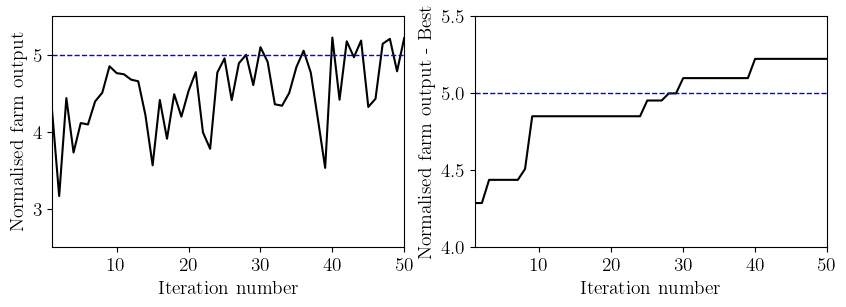}
\caption{Normalised average wind farm power output (left) and best performance history (right).} 
\label{fig:opt_data}
\end{figure}

The optimization runs in batch mode \cite{gonzalez2016batch}, with three sets of parameter values to be explored proposed at every step. This means that convergence is potentially sub-optimal; however, it allows for three numerical experiments to be run in parallel. The optimization is stopped after 50 iterations.  Each simulation required $\approx 70$ CPU hours on ARCHER2. Figure \ref{fig:opt_data} shows the optimization history. The average farm output is normalised by the power produced by a single turbine placed at the center of the available land (denoted $P_0$ and estimated via a separate simulation). The optimization framework succeeds in increasing the power output of the farm from that of the initial layouts, to the point where it exceeds the power that would be produced from five individual turbines by 4.5\%. This is of particular importance, as that would be the maximum output estimated by conventional wake models.

The initial layouts together with two explored during the optimization and the best-performing one are presented in figure \ref{fig:opt_vis}. For each layout, the figure also shows the mean streamwise velocity at turbine hub height. In the case of the best-performing layout (\#40), the turbines are placed side-by-side, minimising wake-turbine interference, with a small streamwise offset so that they can benefit from the local acceleration of the flow around their neighbours. The normalised power produced by each turbine (in ascending order with streamwise distance) and the total normalised farm output are shown in table \ref{tab:power}. Here, the third turbine of layout \#5 is of particular interest, as it is placed at the exact location of the reference turbine, but produces 5\% more power owing to speed-up effects. As before, we note that such an increase in power could not be accounted for by conventional wake models.

\begin{figure}
\adjustboxset{height=.27\textwidth,valign=M}
\begin{tblr}{Q[c,m]|Q[c,m]Q[c,m]}
\hline 
{\bf Case} & {\bf Layout} & $\bar{u}_{h}$ [m/s] \\
\hline \#1 &
\adjincludegraphics{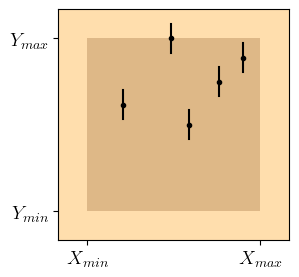} &
\adjincludegraphics{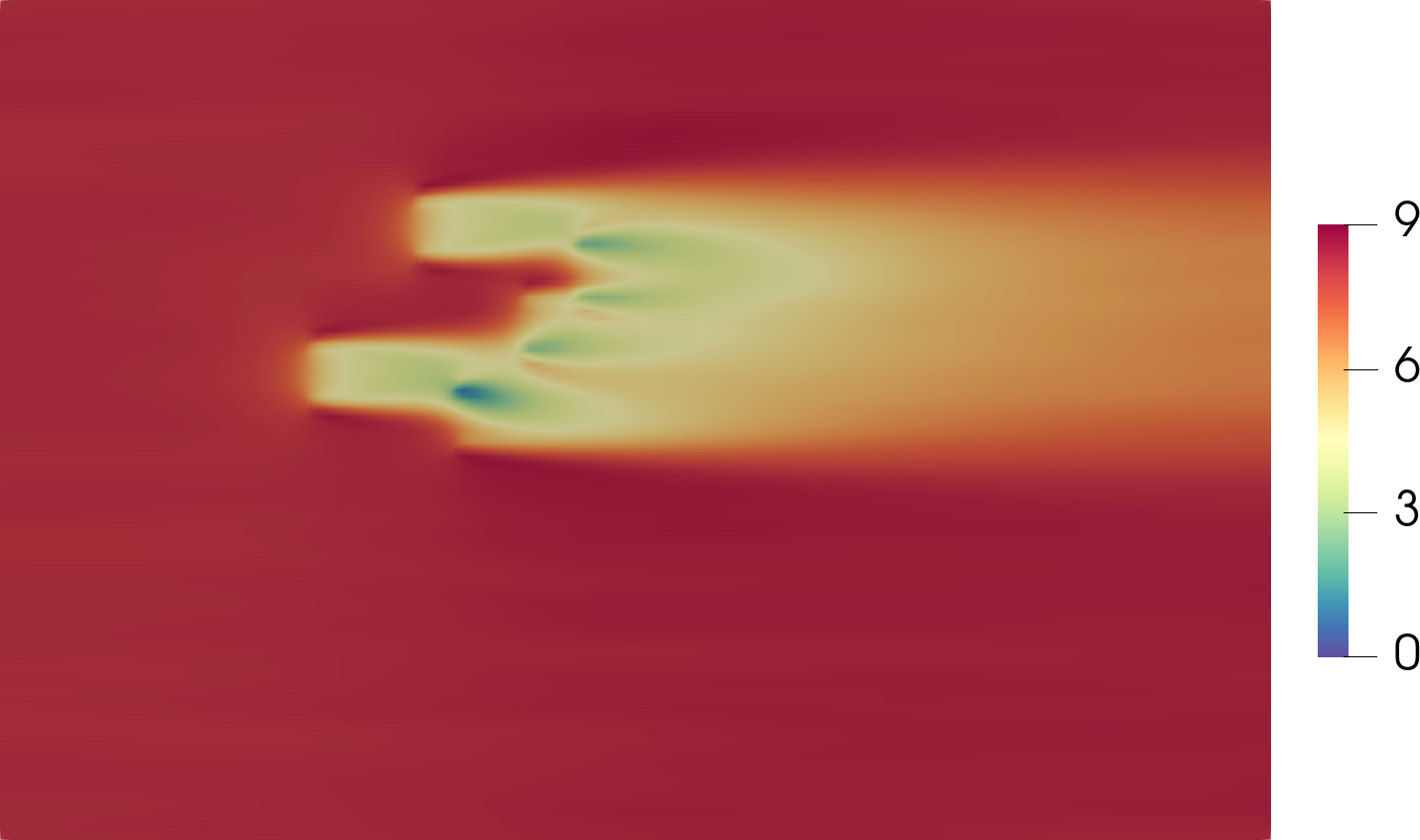} \\
\hline \#2 & 
\adjincludegraphics{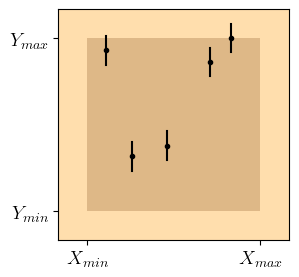} &
\adjincludegraphics{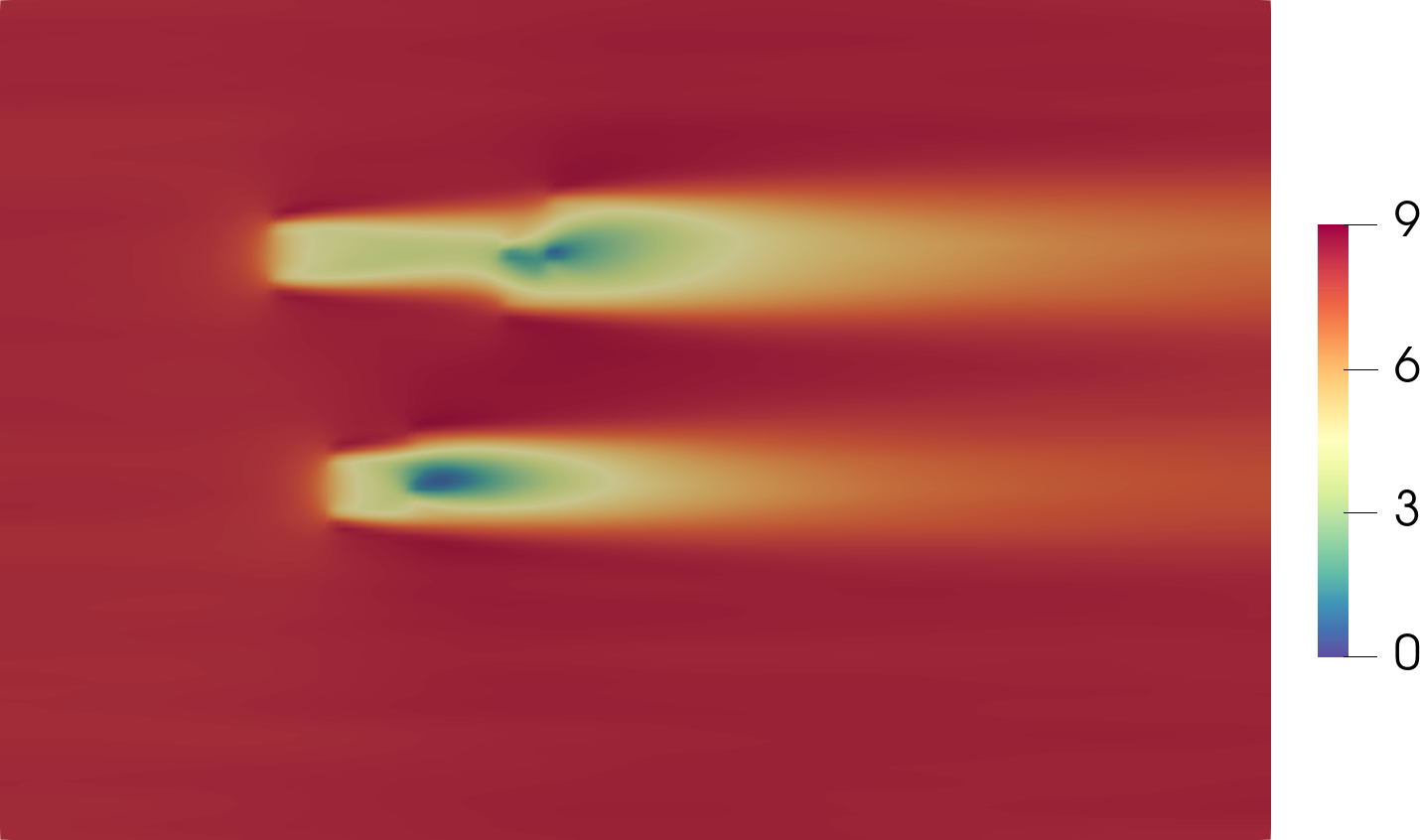} \\
\hline \#3 & 
\adjincludegraphics{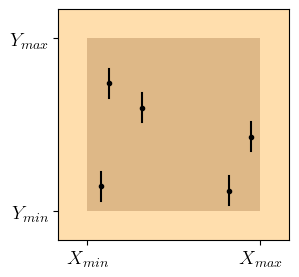} &
\adjincludegraphics{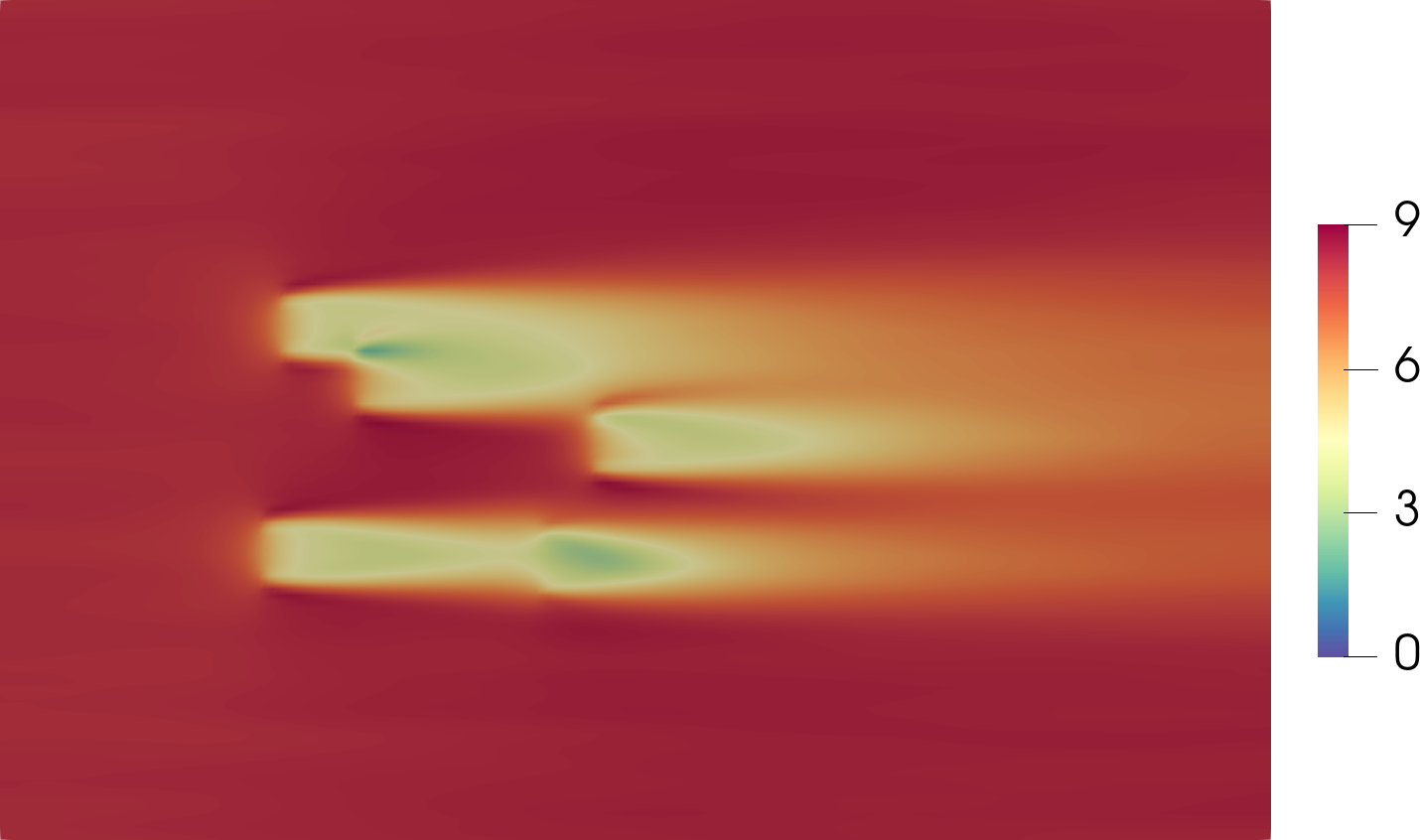} \\
\hline \#4 & 
\adjincludegraphics{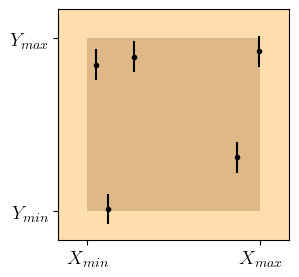} &
\adjincludegraphics{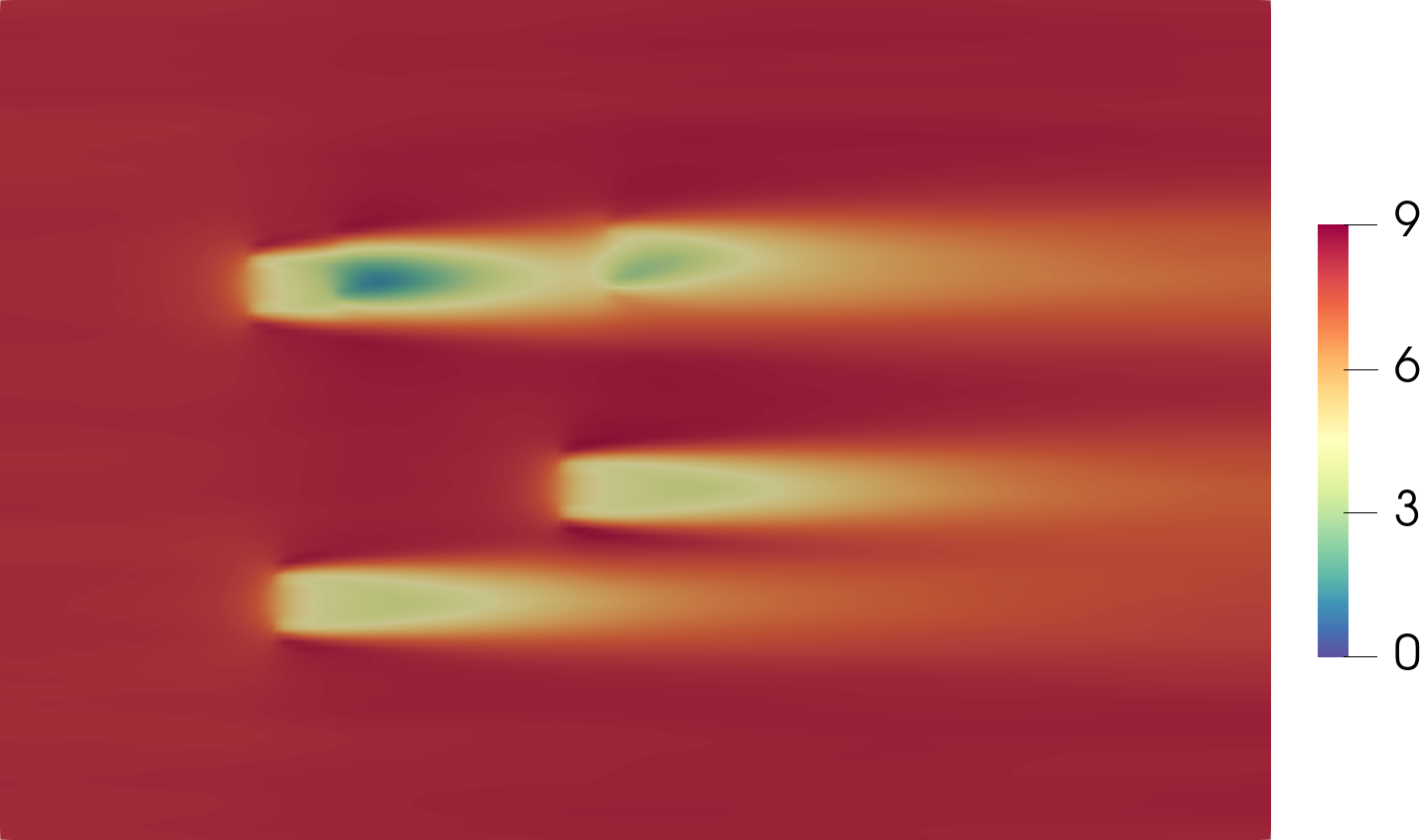} \\
\hline \#5 &
\adjincludegraphics{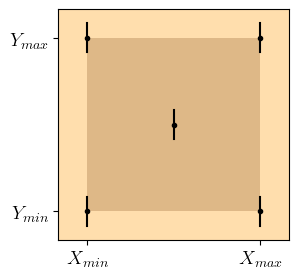} &
\adjincludegraphics{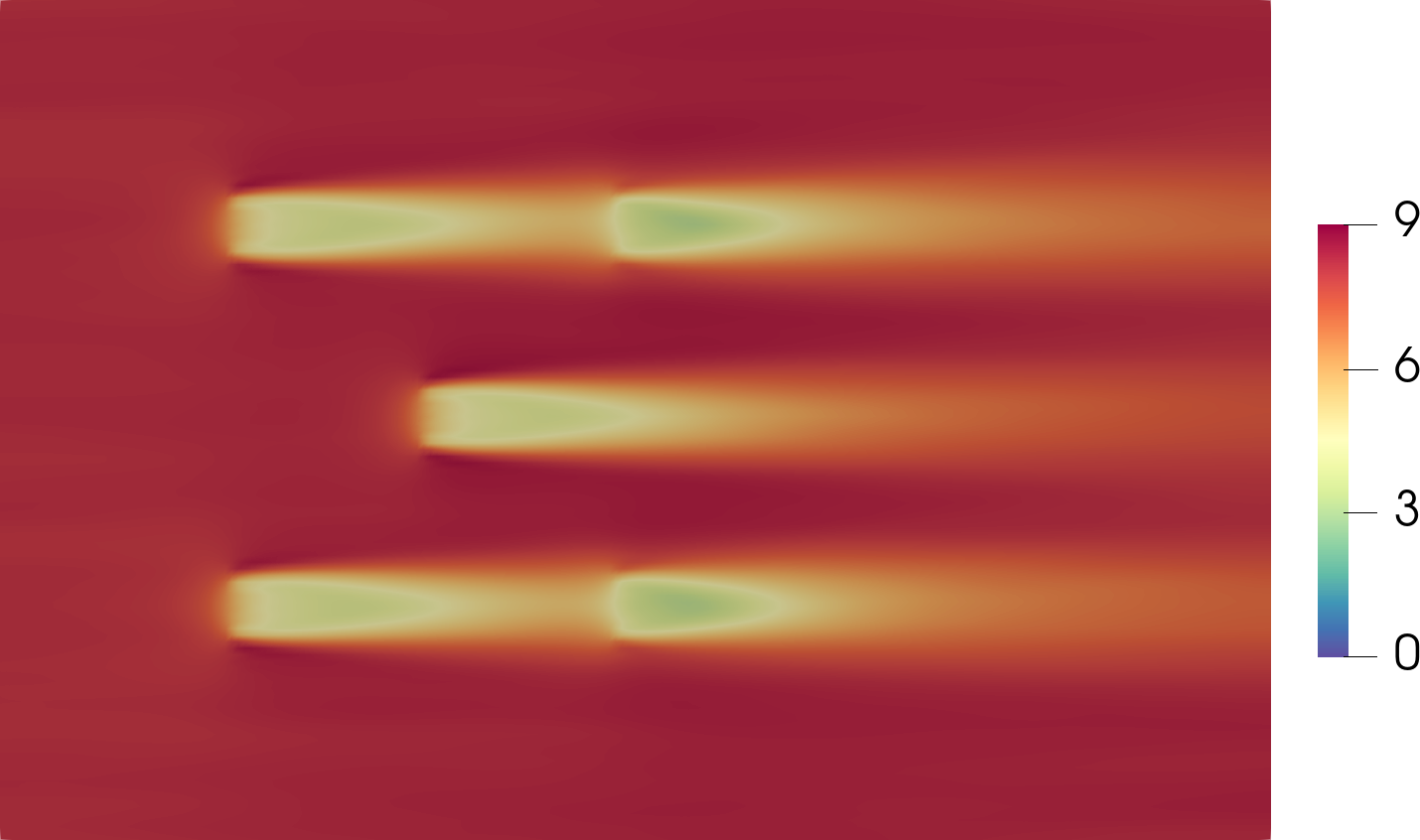} \\
\hline
\end{tblr}
\caption{Example layouts and associated mean streamwise velocity at hub height (continued on the next page).}
\label{fig:opt_vis}
\end{figure}

\begin{figure}
\ContinuedFloat
\adjustboxset{height=.27\textwidth,valign=M}
\begin{tblr}{Q[c,m]|Q[c,m]Q[c,m]}
\hline 
{\bf Case} & {\bf Layout} & $\bar{u}_{h}$ [m/s] \\
\hline \#20 & 
\adjincludegraphics{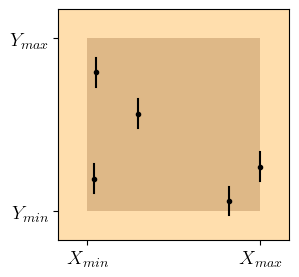} &
\adjincludegraphics{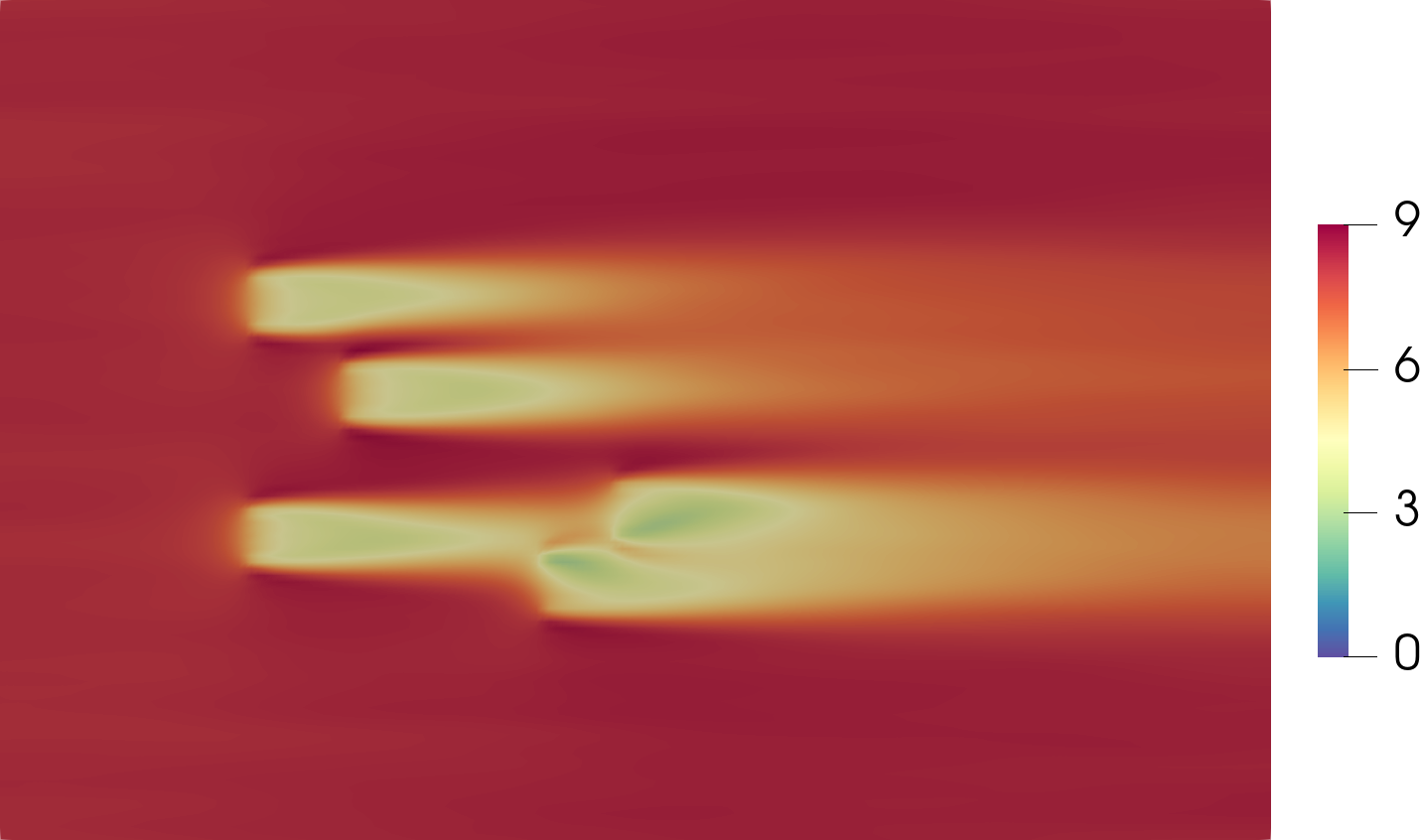} \\
\hline \#30 & 
\adjincludegraphics{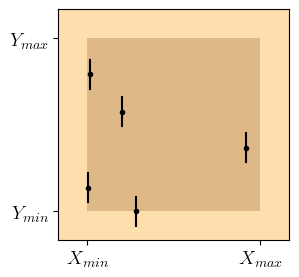} &
\adjincludegraphics{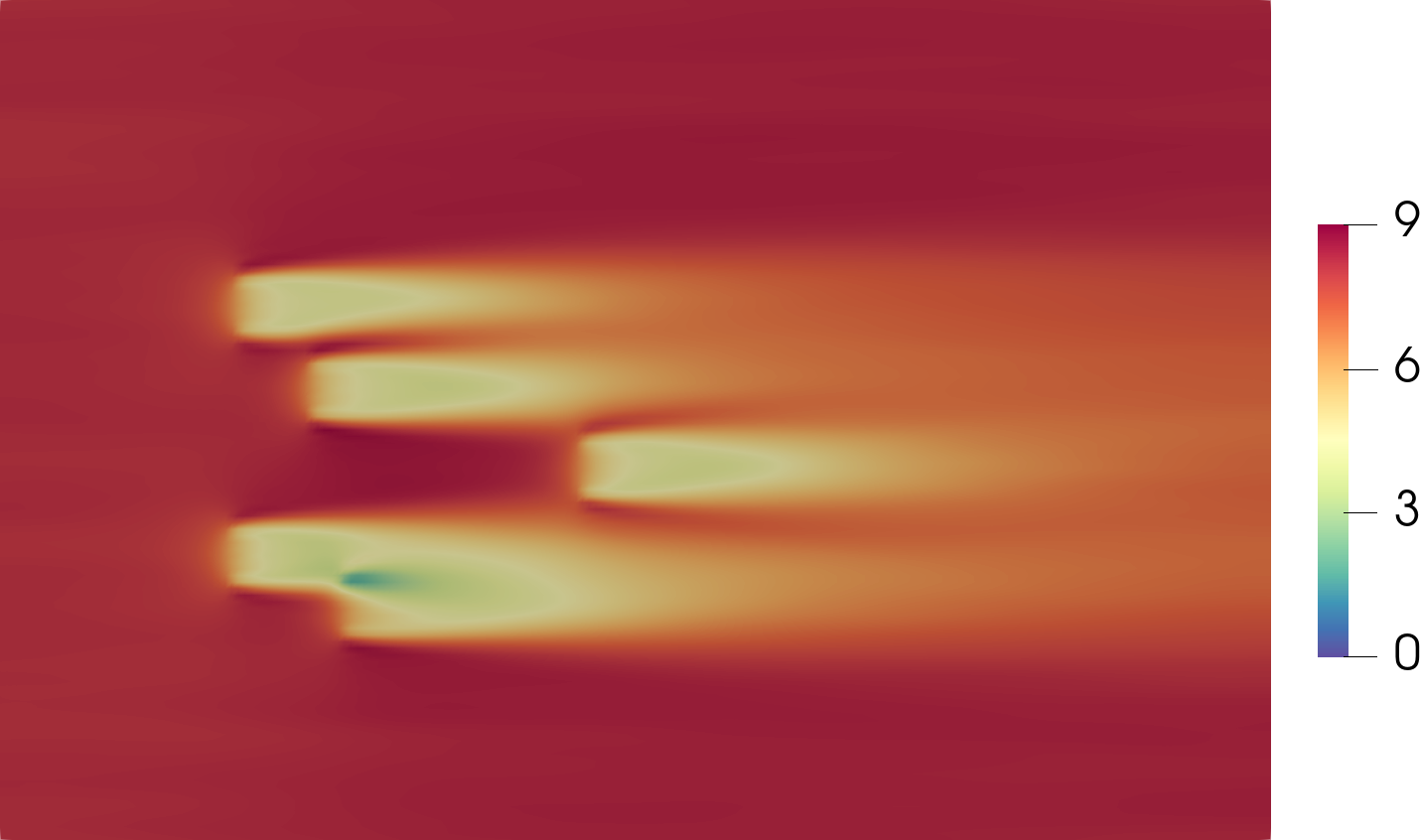} \\
\hline \#40 & 
\adjincludegraphics{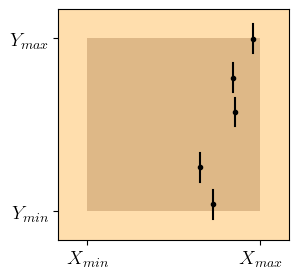} &
\adjincludegraphics{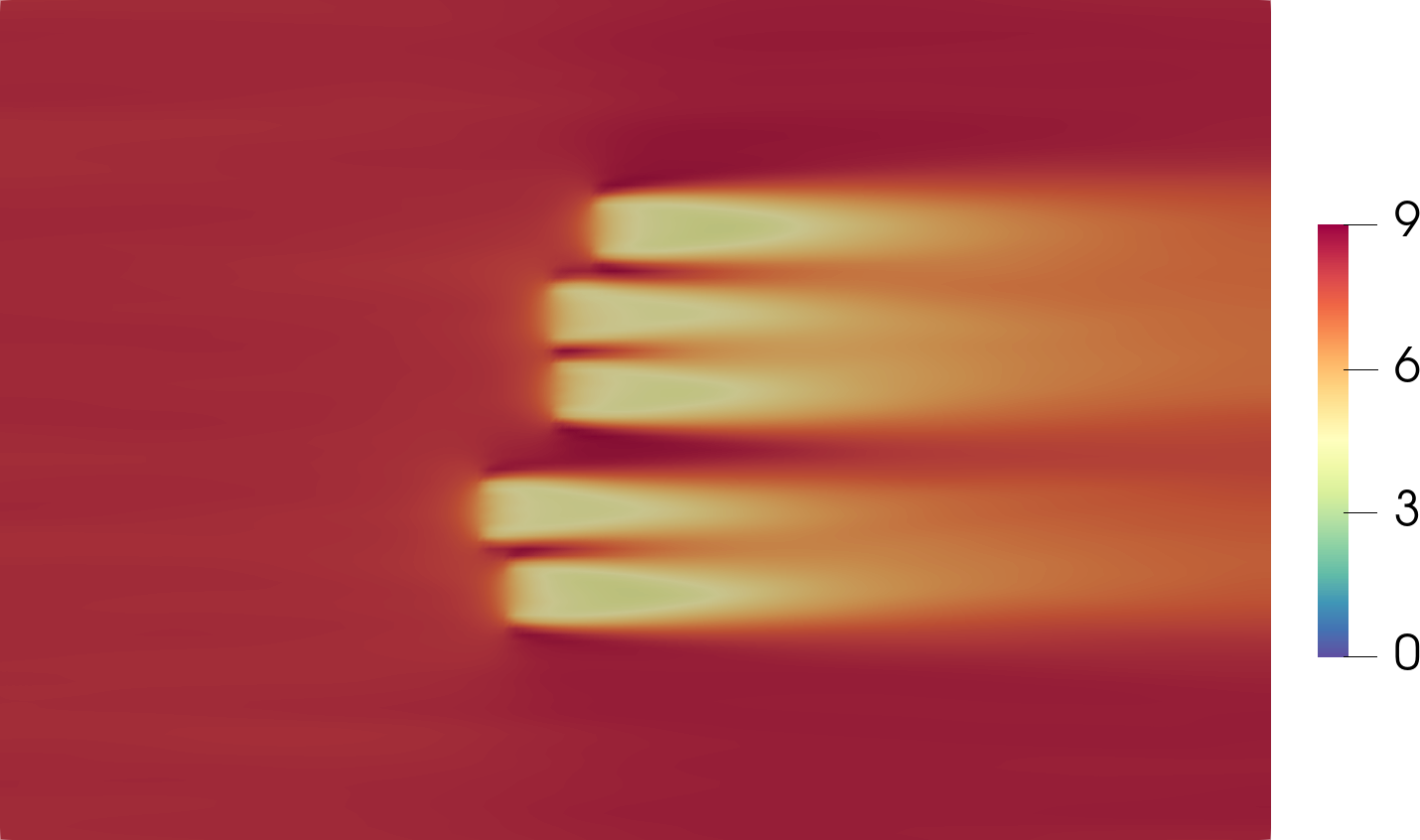} \\
\hline
\end{tblr}
\caption{Example layouts and associated mean streamwise velocity at hub height (continued from the previous page).}
\end{figure}

\begin{table}
\centering
\begin{tblr}{Q[c,m]|Q[c,m]Q[c,m]Q[c,m]Q[c,m]Q[c,m]|Q[c,m]}
{\bf Layout} & $P_1/P_0$ & $P_2/P_0$ & $P_3/P_0$ & $P_4/P_0$ & $P_5/P_0$ & Total \\
\hline 
\#1 & 0.975 & 1.000 & 0.654 & 0.857 & 0.803 & 4.287 \\
\#2 & 0.986 & 0.979 & 0.329 & 0.444 & 0.427 & 3.165 \\
\#3 & 1.005 & 0.987 & 0.989 & 0.398 & 1.059 & 4.438 \\
\#4 & 0.951 & 1.003 & 0.237 & 1.071 & 0.469 & 3.731 \\
\#5 & 1.001 & 1.016 & 1.050 & 0.513 & 0.532 & 4.113 \\
\#20 & 1.006 & 1.003 & 1.083 & 0.806 & 0.634 & 4.531 \\
\#30 & 0.984 & 1.007 & 1.092 & 0.916 & 1.099 & 5.099 \\
\#40 & 1.001 & 1.054 & 1.033 & 1.068 & 1.067 & 5.224
\end{tblr}
\caption{Normalised turbine and farm power outputs for different farm layouts.}
\label{tab:power}
\end{table}

\section{Conclusions}
\label{sec:conclusions}

This study proposes a gradient-free data-driven framework that optimizes the power output of a wind farm using a Bayesian approach and large-eddy simulations of the flow around the farm. Unlike traditional wind farm optimization strategies, which use simple wake models, this framework considers turbulent flow dynamics including wake meandering, wake-to-wake and atmosphere-to-wake interactions. The effectiveness of the framework is demonstrated through a case study of a small wind farm with five turbines. It is shown that the optimization can quickly find optimal designs whilst improving wind farm performance by taking into account these complex flow phenomena.
In the future, the framework will be tested and applied to realistic configurations with complex wind roses (multiple wind directions and velocities) and large wind farms with more turbines. 

\section*{Acknowledgements}
NB and LM are supported by EPSRC, Grant No. EP/W026686/1. LM also acknowledges financial support from the ERC Starting Grant No. PhyCo 949388. The authors would like to thank the UK Turbulence Consortium (EP/R029326/1) for providing access to ARCHER2. 

\bibliographystyle{splncs04}
\bibliography{mybibliography}

\end{document}